\documentclass[aps, 
	prl, 
	noeprint, 
	superscriptaddress, 
	floatfix, 
	tightenlines, 
	twocolumn
	]{revtex4-2}
\usepackage[english]{babel}
\usepackage{graphicx}                      
\usepackage{listings} 
\usepackage{amsmath,amssymb,amsthm}        
\usepackage[utf8]{inputenc}
\usepackage[toc,page]{appendix}
\usepackage{color}
\usepackage{caption}

\usepackage{hyperref}
\hypersetup{
  colorlinks   = true, 
  urlcolor     = blue, 
  linkcolor    = blue, 
  citecolor    = blue 
}

\newcommand{\ket}[1]{\left | #1 \right \rangle}	
\newcommand{\tx}[1]{\textup{#1}}		
\newcommand{\abs}[1]{\left | #1 \right |}		
\newcommand{\proj}[2]{\left | #1 \right \rangle \! \left \langle #2 \right |}	

\begin{document}

\title{Telecom-heralded entanglement between remote multimode solid-state quantum memories}
\author{Dario Lago-Rivera}
\thanks{These two authors contributed equally.}
\author{Samuele Grandi}
\thanks{These two authors contributed equally.}
\author{Jelena V. Rakonjac}
\author{Alessandro Seri}
\affiliation{ICFO - Institut de Ciencies Fotoniques, The Barcelona Institute of Science and Technology, 08860 Castelldefels (Barcelona), Spain.}
\author{Hugues de Riedmatten}
\affiliation{ICFO - Institut de Ciencies Fotoniques, The Barcelona Institute of Science and Technology, 08860 Castelldefels (Barcelona), Spain.}
\affiliation{ICREA - Instituci\'o Catalana de Recerca i Estudis Avan\c cats, 08015 Barcelona, Spain.}

\begin{abstract}
Future quantum networks will enable the distribution of entanglement between distant locations and allow applications in quantum communication, quantum sensing and distributed quantum computation. At the core of this network lies the ability of generating and storing entanglement at remote, interconnected quantum nodes. While remote physical systems of various nature have been successfully entangled, none of these realisations encompassed all of the requirements for network operation, such as telecom-compatibility and multimode operation.
Here we report the demonstration of heralded entanglement between two spatially separated quantum nodes, where the entanglement is stored in multimode solid-state quantum memories. At each node a praseodymium-doped crystal stores a photon of a correlated pair, with the second photon at telecommunication wavelengths. Entanglement between quantum memories placed in different labs is heralded by the detection of a telecom photon at a rate up to 1.4 kHz and is stored in the crystals for a pre-determined storage time up to 25 microseconds. We also show that the generated entanglement is robust against loss in the heralding path, and demonstrate temporally multiplexed operation, with 62 temporal modes. 
Our realisation is extendable to entanglement over longer distances and provides a viable route towards field-deployed, multiplexed quantum repeaters based on solid-state resources.
\end{abstract}

\maketitle

Fast developing quantum simulation and computation centres, as well as secure communication systems, will soon require a reliable network for the distribution of entanglement. A promising blueprint for such a network, based on the quantum repeater architecture \cite{Duan2001,Simon2007}, relies on distributing entanglement in a heralded fashion between remote quantum nodes, where information can be stored and manipulated in stationary, matter qubits. Entanglement between separate material systems has been demonstrated with atomic ensembles \cite{Chou2005,Chou2007,Yuan2008,Yu2020}, single trapped ions and atoms \cite{Moehring2007,Ritter2012,Hofmann2012} and more recently with solid-state systems \cite{Bernien2013,Usmani2012, Stockill2017,Sipahigil2016,Humphreys2018}. However, integration into a quantum network will require a high entanglement rate with heralded operation, compatibility with the telecom network, long storage times and particularly multimodality, a combination that hasn't been achieved yet.

Among possible candidates, rare-earth doped solids played a pivotal role in the development of quantum memories \cite{deRiedmatten2008}, providing a system with a large number of atoms naturally trapped in a solid-state matrix and with excellent coherence properties.
Fundamental progress has been reported in rare-earth based systems, including high-efficiency \cite{Hedges2010} and long-lived \cite{Laplane2017} quantum storage, as well as light-matter entanglement \cite{Clausen2011,Saglamyurek2011,Ferguson2016,Kutluer2019}.
A fundamental advantage of rare-earth based quantum memories is the large multiplexing capability, combining temporal, spectral and spatial degrees of freedom \cite{Afzelius2009,Sinclair2014,Seri2019,Yang2018}. While entanglement between two such memories has already been demonstrated, it was either in a non heralded fashion \cite{Puigibert2020} or with a technique not directly extendable to long distance communication \cite{Usmani2012}.

Here we move forward towards the realisation of the unitary element of a quantum network by demonstrating scalable, telecom-heralded matter-matter entanglement between two remote multimode solid-state quantum memories. We demonstrate a heralding rate in the kHz regime and a storage time up to 25 $\mu s$, resulting in the storage of 62 temporal modes. We also show that our implementation is resistant against loss, and directly extendable to operation over long-distances.

\begin{figure*}[t]
	\centering
	\includegraphics[width = \textwidth]{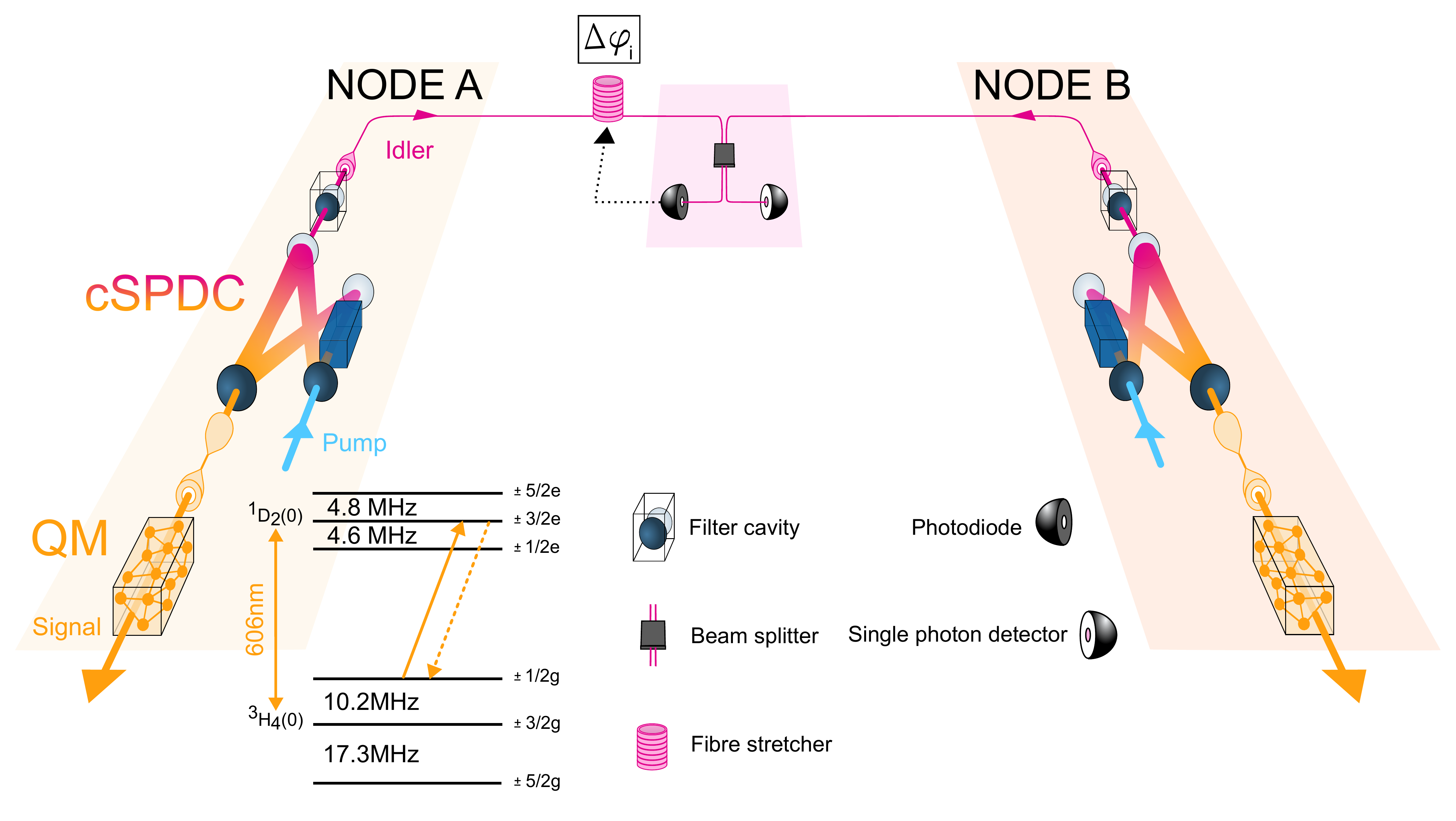}
	\caption{Schematics of the experiment. Each node hosts a cSPDC and a QM, with the two memories placed in two labs 10 m apart, and separated by 50 m of fibre. Each signal photon is fibre coupled and sent to a Pr-doped memory, where it is stored through the AFC protocol. Fibre-Coupled idler photons are mixed at a fibre beam-splitter in the central station. Here they are detected with superconducting single photon detectors, where a click heralds the generation of an entangled state of the two quantum memories. One of the output of the fibre beam-splitter is connected to a photodiode, used to lock the phase of the idler paths. The inset shows the energy levels of Pr, where is highlighted the transition where the AFC is prepared.}
	\label{fig1}
\end{figure*}
Our proposal, sketched in Fig.~\ref{fig1}, is a hybrid approach, in which a single collective atomic excitation, created by non-degenerate parametric down-conversion sources, is shared between two remote Praseodymium (Pr) quantum memories (QM) \cite{Simon2007}. The two sources are based on type-I cavity-enhanced spontaneous parametric down-conversion (cSPDC) that, from a pump laser at 426 nm, generates photon pairs: a telecom idler at 1436 nm and a signal at 606 nm, in resonance with the $^3$H$_4$ $\leftrightarrow$ $^1$D$_2$ transition of Pr$^{3+}$:Y$_2$SiO$_5$ \cite{Seri2018}. The source cavity allows for a bi-photon linewidth of 1.8 MHz, matching the memory bandwidth. To ensure joint emission of indistinguishable photon pairs the relative length of the two cavities, each $\sim$1.2 m long, was adjusted to within a few nm \cite{SuppMat}.
Two filter cavities selected the central frequency modes of the idler channels. Both idler and signal photons were coupled in single mode fibres, with the latter sent to different quantum memories.
These are two Pr-doped Y$_2$SiO$_5$ crystals that are located in two closed-cycle cryostats, placed in different laboratories, roughly 10 m apart. Quantum light is stored in the crystals through the Atomic Frequency Comb (AFC) protocol \cite{Afzelius2009}. The inhomogeneously broadened profile of the Pr-ion ensemble is tailored into a comb-like structure on the $\pm1/2g \rightarrow \pm 3/2e$ transition by means of optical pumping. Light absorbed by the memory will be re-emitted after a fixed time $\tau = 1/\Delta$, where $\Delta$ is the period of the frequency comb. The signal photons were focused into the crystal to a waist of 40 $\mu$m, where they were stored as collective excitations $\sum_k^N c_k \ket{g_1 \ldots e_k \ldots g_N}$ of ground $\ket{g}$ and excited $\ket{e}$ states of about $N = 10^{7}$ Pr-ions.

Entanglement between the two quantum memories is achieved by mixing the two idler modes on a fibre beam-splitter, thus erasing the which-path information. A detection event after the idler beam-splitter heralds, in the ideal case, the generation of the entangled state
\begin{equation}
	\ket{\psi} = \frac{1}{\sqrt{2}} \left ( \ket{0}_\tx{A} \ket{1}_\tx{B}  + e^{\Delta \varphi_i} \ket{1}_\tx{A}  \ket{0}_\tx{B}  \right )
\end{equation}
where 0 and 1 represents the number of delocalised atomic excitations mapped in the two crystals at nodes A and B. $\Delta \varphi_i$ is the relative phase between the two idler channels at the beam splitter, that has to be locked for consistent heralding. Losses in the signal path, multi-photon generation events, phase noise in the idler path as well as limited detection and heralding efficiencies would turn $\ket{\psi}$ into a mixed state. It is therefore crucial to operate in the single excitation regime, and to effectively control $\Delta \varphi_i$. The entanglement between the two nodes can then be estimated through a tomography of the joint state \cite{Chou2005}.
\begin{figure*}[t]
	\centering
	\includegraphics[width = 1.5 \columnwidth]{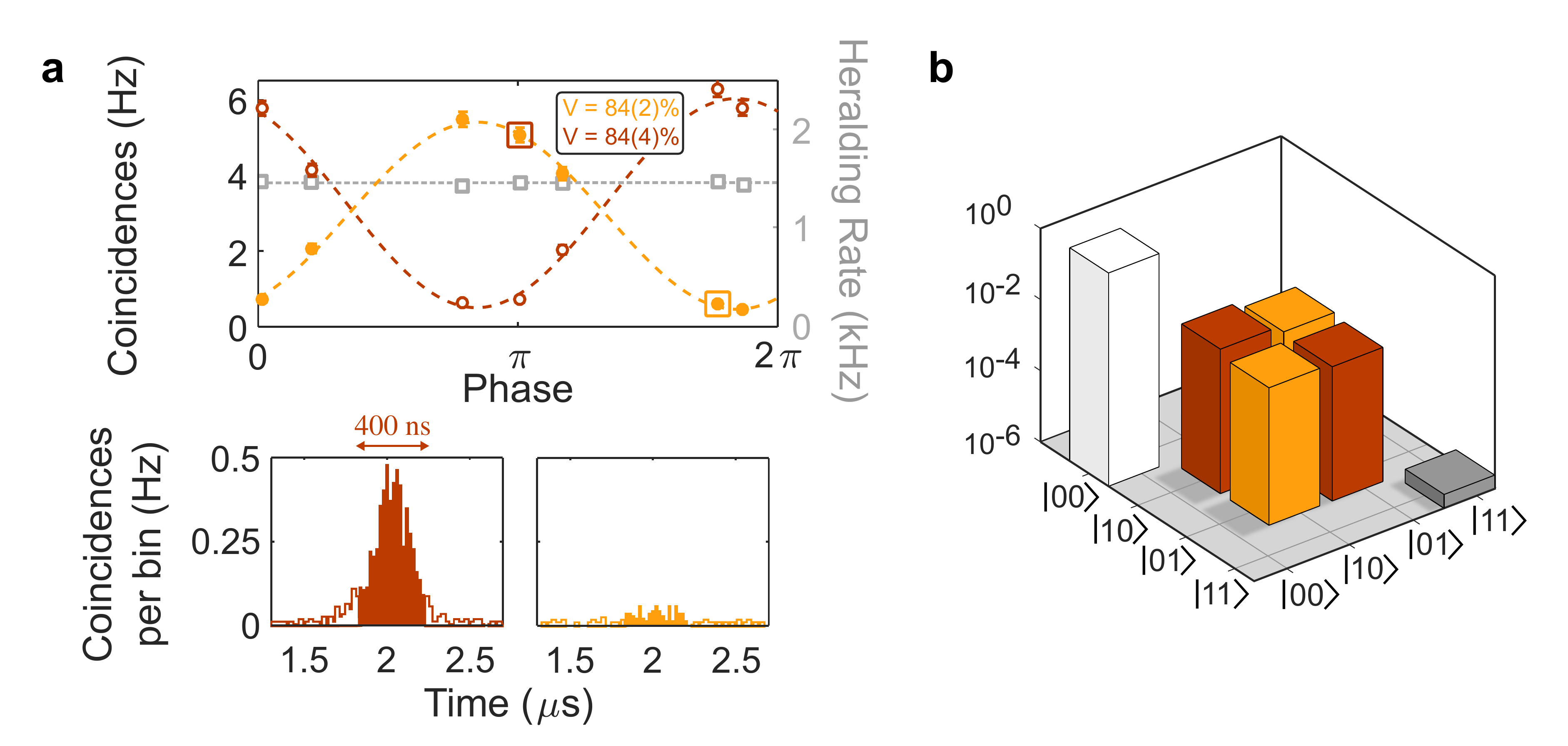}
	\caption{Entanglement verification for 2 $\mu$s AFC. (a) Interference fringes obtained at the two output of the beam-splitter by recombining the two signal modes (circles) and heralding rate (squares). Each point was obtained integrating for about 2 minutes, and with a coincidence window of 400 ns. The two bottom plots represent the actual coincidence histograms for the points indicated in the fringe plot. (b) Reconstructed density matrix for the matter-matter entangled state of the two remote quantum memories. The measurement of the diagonal elements took 90 minutes.}
	\label{fig2}
\end{figure*}

We first demonstrated matter-matter entanglement between the two remote quantum memories, with 2 $\mu$s of storage time. We used an intermittent locking system to stabilise the setup, where $\Delta \varphi_i$ was controlled adjusting a piezo-electric fibre stretcher \cite{Seri2018,SuppMat}. We prepared the AFCs of both memories such that the storage times and temporal widths of the echoes would be equal, and we synchronised the preparation of the two AFCs to match the two storage temporal windows.
We measured an entanglement heralding rate of 1.43 kHz at one output of the idler beam-splitter, with a duty cycle of 43 $\%$. This value, limited by the available pump power \cite{SuppMat}, is about 40x higher than the current record of distribution between long-lived memories, albeit with a lower quantum link efficiency \cite{Humphreys2018}.

We quantified the matter-matter entanglement between the two nodes by mapping the atomic excitations back to photons and measuring the concurrence $\mathcal{C}$ of the two modes \cite{Chou2005}. This can be expressed as $\mathcal{C} = \mathtt{max} \left [ 0, \, \left ( 2 \abs{d} - 2 \sqrt{p_{00} \,  p_{11}} \right ) \right ]$, where $p_{ij}$ are the probabilities of having $i(j)$ excitations in mode A(B), $d = V \left ( p_{01} + p_{10} \right )/2$ and $V$ is the visibility of the interference between the two modes. While the different probabilities $p_{ij}$ can be inferred from photon statistics, it is necessary to mix the two signal paths on a beam-splitter to measure $V$. To stabilise this interferometer, with a total fibre length close to 75 m, we employed another intermittent locking scheme where we periodically injected light detuned from the AFC. This allowed us to stabilise the relative phase between the two signal paths using a piezo-electric fibre stretcher \cite{SuppMat}, while leaving the QM undisturbed.
The results are summarised in Fig.~\ref{fig2}. We measured a high suppression of two-photon events \cite{Chou2005}, $h^{(2)}_c = p_{11} / (p_{10} p_{01}) = 0.036(8)$, and interference visibilities of 84(2)$\%$ and 84(4)$\%$ at the two outputs of the signal beam-splitter. This accounted to a total concurrence of $\mathcal{C}_1 = 1.15(5) \cdot 10^{-2}$ and $\mathcal{C}_2 = 1.15(7) \cdot 10^{-2}$. As the AFC emission is a local operation, and both values of $\mathcal{C}$ are above zero by more than 20 and 15 standard deviations respectively, this is a definite confirmation that the two crystals shared an entangled state.
\begin{figure*}[t]
	\centering
	\includegraphics[width = 1.8 \columnwidth]{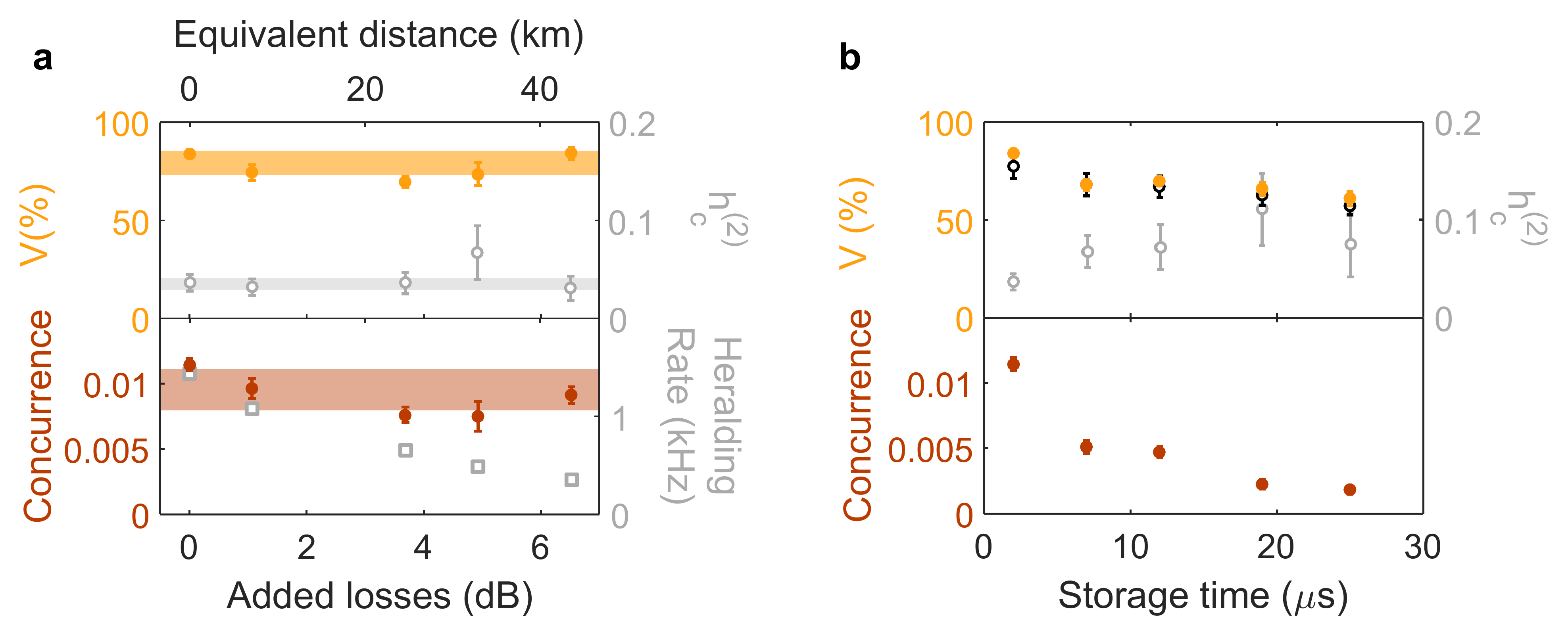}
	\caption{Concurrence for different experimental configurations. (a) Increasing losses in the idler channels. While the heralding rate decreases, the concurrence remains constant. 
	The solid bands represent a sigma variation from the mean values of visibility, $h^{(2)}_c$ and concurrence. (b) Increasing storage time. We increased $\tau$ up to 25 $\mu$s, and measured positive concurrence for all cases. 
	The empty dark circles are the expected values for visibilities.}
	\label{fig3}
\end{figure*}
The high contribution of $p_{00}$ to the density matrix, visible in inset (b) of Fig.~\ref{fig2}, is mostly due to losses in the signal channels. It can be reduced by back-tracing the value of $\mathcal{C}$ to the crystal: the concurrence is then rescaled to $\widetilde{\mathcal{C}}_1 = 7.3(5) \cdot 10^{-2}$ and $\widetilde{\mathcal{C}}_2 = 7.4(5) \cdot 10^{-2}$ \cite{SuppMat}. Moreover, in a Duan-Lukin-Cirac-Zoller like quantum repeater \cite{Duan2001,Chou2007} with built-in purification the $p_{00}$ term would only affect the entanglement distribution rate but not the fidelity of the repeater. To provide an estimation of how well our entangled link would perform in such a repeater we defined an effective fidelity considering only the one- and two-photons terms of the matrix \cite{SuppMat}, and we inferred values of $\mathcal{F}^\tx{eff}_1 = 0.92(1)$ and $\mathcal{F}^\tx{eff}_2 = 0.92(2)$.

This manner of distributing entanglement using single-photon detection is particularly suited for operation over long distances \cite{Simon2007,Sinclair2014,Yu2020}. The two nodes, each containing a source and a quantum memory, could be placed several kilometres apart, with the idler mixing station placed between them and connected by telecom fibres. Nevertheless, the additional propagation losses would not affect the concurrence, but only lower the heralding rate. To verify this statement we used variable attenuators to add losses in both idler channels, to mimic the increased attenuation in long optical fibres. We measured the concurrence for losses up to 6.5 dB per channel: as shown in Fig.~\ref{fig3}(a), the concurrence does not change appreciably. The point-to-point fluctuation is due to the change in visibility of the interference between the single excitation states, that are affected by imbalances between the two sources \cite{SuppMat}; as expected, $h^{(2)}_c$ is instead constant.

Even if the propagation losses would not affect the quality of the entanglement distribution, in a real demonstration a greater distance between the nodes would require a longer storage time in the QMs, to allow the idler photons to reach the central station and for the heralding signal to come back to the node - the so-called communication time $t_\tx{com}$. We therefore increased $\tau$ for the two QMs, and measured again the entanglement.
As reported in Fig.~\ref{fig3}(b), the concurrence decreased with increasing $\tau$ due to the reduced efficiency of the storage protocol, resulting in a higher contribution of $p_{00}$. The visibility also decreased, due to decreasing overlap between the AFC echoes at longer storage times \cite{SuppMat}. Nevertheless, even for the longest $\tau$ of 25 $\mu$s, about $\sim10^3$ longer than previous demonstrations in solid-state rare-earth quantum memories \cite{Usmani2012,Puigibert2020}, the positivity of the concurrence was maintained with more than 5 standard deviations, allowing heralded entanglement over 5 km of optical fibre.

\begin{figure*}[t]
	\centering
	\includegraphics[width = 2 \columnwidth]{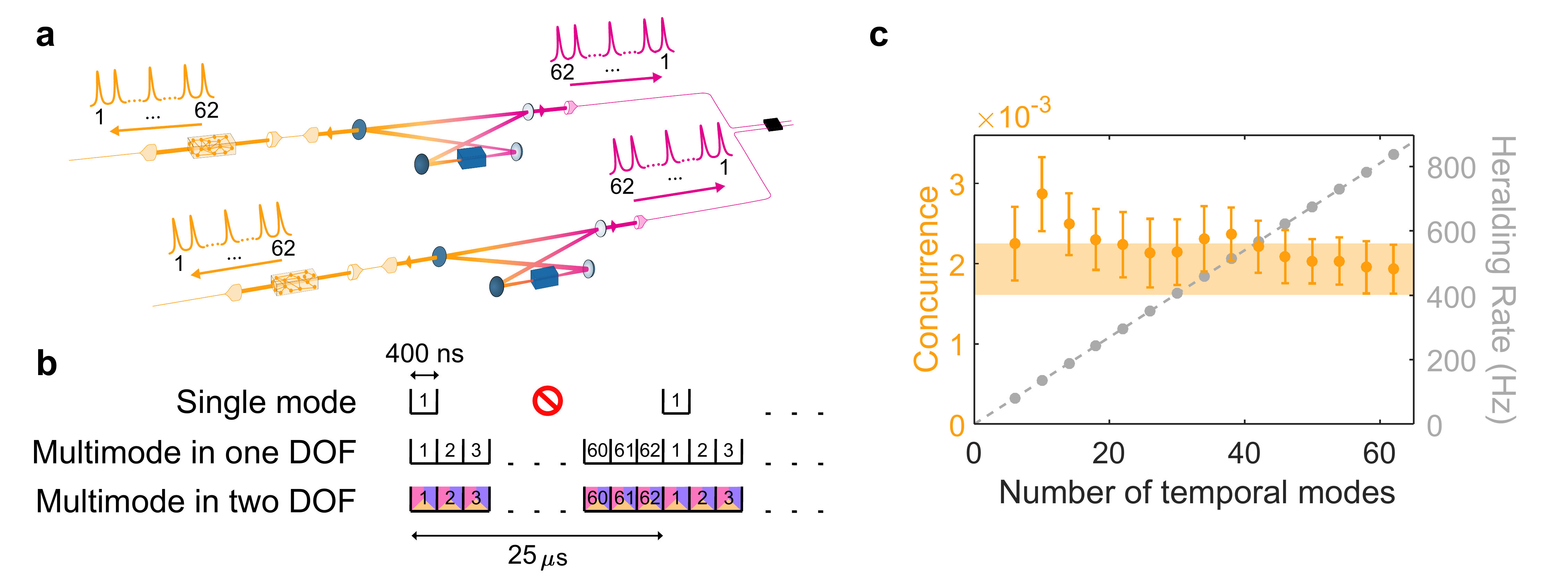}
	\caption{Multimode operation of a quantum memory. (a)
	temporally multimode operation, where a train of pulses is absorbed by both memories, and re-emitted in the same order. (b) advantage of multimode memories over single mode memories, as modes can be stored during the whole communication time. DOF: degrees-of-freedom. (c) concurrence and heralding rate for increasing number of modes. The solid band represent a sigma variation from the mean value with 62 modes. The variation of the concurrence trace is dictated mostly from the uncertainty in the visibility. The heralding rate increases linearly with the number of modes considered. It is lower than recorded for lower storage times due to the lower duty-cycle.}
	\label{fig4}
\end{figure*}

Finally, we would like to emphasise the advantage that comes from the multimodality of this protocol. Thanks to the nature of the AFC storage, several temporal modes can be stored in the QM before the first absorbed mode has been re-emitted \cite{deRiedmatten2008}. This means that it is not necessary to wait for the entanglement heralding signal from the central station before storing another mode. Multimode operation then becomes particularly useful for operation over long distances, where $t_\tx{com}$ seriously limits the entanglement distribution rate for single-mode memories.
To investigate this effect we analysed the data acquired for the longest storage, where we considered a $t_\tx{com}$ of 25 $\mu$s, equal to the AFC storage and simulating crystals 5 km apart. Considering a single temporal mode duration of 400 ns, that comprises $90 \%$ of the photon, we split $\tau$ into 62 temporal modes. By dividing the whole measuring time in ``communication trials" of 25 $\mu$s, we sorted the idlers into the 62 modes according to their arrival time. We were then able to calculate the concurrence and heralding rate for an increasing number of allowed modes \cite{SuppMat}. The results are reported in Fig.~\ref{fig4}(c).
While the concurrence remained constant, the heralding rate increased linearly with the number of modes allowed. This increase becomes more and more dramatic as longer distances and longer $t_\tx{com}$ are considered.

The remote matter-matter entanglement demonstrated in this work represents a crucial step towards a viable realisation of a quantum repeater, showing for the first time telecom-heralded entanglement between two multimode quantum memories. Our heralding rate was limited by the available cSPDC pump power, and could be boosted to $\sim$15.6 kHz while maintaining positive concurrence \cite{SuppMat}. We showed that our scheme is extendable to long distances, and integration into field-deployed quantum networks could soon be achieved thanks to the depth of possibilities available to these systems.
First, entanglement between fully decoupled setups could be realised by employing a remote verification scheme \cite{Caspar2020}. Longer distances could be bridged with spin-wave operation, leading to longer storage times and on-demand retrieval \cite{Simon2007,Seri2017,Laplane2017}, and offering a synchronisation tool to extend the entanglement across several links. A different approach would entail greatly increasing the multimodality for a fixed storage time operation \cite{Sinclair2014}. Indeed, a potential for thousands of stored modes has been envisioned for these memories by harnessing its available temporal, frequency and spatial multimodality \cite{Seri2018,Seri2019,Yang2018}. Additionally, the solid-state nature of these quantum memories opens the way to chip integration \cite{Saglamyurek2011,Zhong2017,Seri2018}.

\section*{Acknowledgments}
This project received funding from the European Union Horizon 2020 research and innovation program
within the Flagship on Quantum Technologies through grant 820445 (QIA) and under the Marie Sk\l odowska-Curie grant agreement No. 713729 and No. 758461, by the Gordon and Betty
Moore foundation through Grant GBMF7446 to HdR, by the Government of Spain (PID2019-106850RB-I00; Severo Ochoa CEX2019-000910-S; BES-2017-082464), Fundaci\'o Cellex, Fundaci\'o Mir-Puig, and Generalitat de Catalunya (CERCA, AGAUR).

\clearpage
\section*{Appendices}
\appendix
\section{Phase locking}
The relative phase between the idler photons interfering at the fibre BS in the central station needs to be locked to achieve consistent heralding. In addition, since the entanglement of the heralded state is verified by recombining the signal photons on another BS, their relative phase has to be controlled as well.

We achieved it by alternating between stages of locking and measuring. During the locking stage, we injected classical fields in the setup as a reference. To lock the relative phase between the idler modes we used a classical beam at 1436 nm. This field is naturally generated by a difference-frequency-generation process in the same PPLN crystals that we use for the cavity-enhanced photon pair generation. It is a valid reference as it has the same wavelength of the idler photons and it shares the same optical path. In the case of the signal photons, we used light 300 MHz detuned from the AFC frequency to avoid disturbing the comb spectrum. See Supplementary Materials for additional information.

For both idler and signal stabilisation, we detected the interference between the classical reference fields at the two BSs with a photodiode, which we use as the input of a PID controller. We feedbacked on home-made fibre stretchers that we built by rolling a portion of the optical fibres around cylindrical piezoelectric actuators.

\section{Limit on the fringe visibilities}
There are three main factors that limited the visibilities that we measured and that we account for in Fig.~\ref{fig3}(b) of the main text. 
\begin{description}
	\item[Phase noise in optical fibres] as mentioned in the previous section, we alternated between locking stages and measuring stages. During the latter there is no control over the phase evolution in the fibres, as the classical reference fields are blocked. Despite the good passive stability that we achieved in our fibres, the contribution of phase noise in the fibres carrying the signal photons is still considerable and is the main factor that limits the visibility that we measured.
	\item[Indistinguishability of the idler modes] we quantified the overlap between the idler modes generated by the two sources by performing a Hong-Ou-Mandel experiment. After correcting for accidental counts we estimated an overlap of $\eta_\tx{ov} = 90(7) \%$.
	\item[Overlap of the retrieved signal modes] the state tomography that we employed required a recombination of the signal modes after storage in the AFC. Therefore, we adjusted the two AFCs in order to best match the wave-packets of the two retrieved signal photons. However, because of the different experimental conditions of the two memories, placed in different laboratories and prepared from different setups, this overlap was not perfect. For each case we measured the intensity, width and position of each echo, and calculated the maximum interference visibility with these parameters.
\end{description}
The final expected visibility was calculated as the product of these three factors.

\section{Effective fidelity}
The density matrix of the reconstructed entangled state has a high $p_{00}$ contribution, as visibile in Fig.~\ref{fig2}(b) of the main text. In a DLCZ-like quantum repeater \cite{Duan2001} the vacuum components will affect the rate of the experiment, but not the fidelity of the distributed entanglement. This is due to the fact that the DLCZ scheme includes built-in purification by using two chains of entangled memories. 

In this case only events with one photon at each side will be considered, therefore removing the vacuum component of the density matrix. We then infer the effective fidelity $\mathcal{F}^\tx{eff}$ as $\left ( \mathtt{Tr} \sqrt{\sqrt{\widetilde{\rho}}\, \sigma \sqrt{\widetilde{\rho}}} \right )^2$, where $\widetilde{\rho}$ is the reconstructed state where we fix the $\proj{00}{00}$ component to zero, and $\sigma$ is the ideal state $\proj{\psi}{\psi} = 1/2 \left ( \proj{10}{10} + \proj{10}{01} + \proj{01}{10} + \proj{01}{01} \right )$. In the notation with $p_{ij}$, the fidelity can be expressed as:
\begin{equation}
	\mathcal{F}^\tx{eff} = \frac{1}{2} \frac{(\tilde{p}_{01} + \tilde{p}_{10})(1 + V)}{\tilde{p}_{01} + \tilde{p}_{10} + \tilde{p}_{11}}
\end{equation}
The fidelity reported in the main text was calculated using values of $\tilde{p}_{ij}$ backtraced to the crystal, and the error was calculated with standard error propagation.

\bibliography{qpsa}

\end{document}